\documentclass{elsart4}

\usepackage{amssymb}
\usepackage{graphicx}
\usepackage{epsfig}
\usepackage{color}
\usepackage[english,francais]{babel}
\usepackage{bm}

\newtheorem{e-proposition}[theorem]{Proposition}

\newtheorem{e-definition}[theorem]{Definition\rm}

\def\og{\leavevmode\raise.3ex\hbox{$\scriptscriptstyle\langle\!\langle$~}}
\def\fg{\leavevmode\raise.3ex\hbox{~$\!\scriptscriptstyle\,\rangle\!\rangle$}}

\newcommand{\beq}{\begin{equation}}
\newcommand{\eeq}{\end{equation}}
\newcommand{\ba}{\begin{eqnarray}}
\newcommand{\ea}{\end{eqnarray}}

\begin{document}
% Select a primary header Physics or Astrophysics
% You can place after the header (classification), if you know it.

%\centerline{Physics or Header}
\centerline{Mechanics or Header}
%\begin{frontmatter}


% \author{Name\thanksref{label2}}
\selectlanguage{english}
\title{On a class of three-phase checkerboards with unusual effective properties}

% If all authors are at the same address, the [label1] can be suppressed

\selectlanguage{english}
\author[authorlabel1]{Richard V. Craster}
\ead{r.craster@imperial.ac.uk}
\author[authorlabel2]{S\'ebastien Guenneau}
\ead{sebastien.guenneau@fresnel.fr}
\author[authorlabel3]{Julius Kaplunov}
\ead{julius.kaplunov@brunel.ac.uk}
\author[authorlabel3]{Evgeniya Nolde}
\ead{evgeniya.nolde@brunel.ac.uk}

\address[authorlabel1]{Department of Mathematics, Imperial College
London, London SW7-2AZ, UK}
\address[authorlabel2]{Institut Fresnel, UMR CNRS 6133, University of Aix-Marseille, France}
\address[authorlabel3]{Department of Mathematical Sciences, Brunel University, Uxbridge, Middlesex, UB8 3PH, U.K.}

% If your know the dates of reception, and acceptation you can put them now;
%    idem the name of the person presenting your article

\medskip
%\begin{center}
%{\small Received *****; accepted after revision +++++}
%\end{center}

%\begin{abstract}
%\end{abstract}
\noindent{\bf Abstract}
\vskip
0.5\baselineskip
We examine the band spectrum, and associated Floquet-Bloch
eigensolutions, arising in a class of three-phase periodic checkerboards.
On a periodic cell $[-1,1[^2$, the refractive index is defined by $n^2= 1+ g_1(x_1)+g_2(x_2)$
with $g_i(x_i)=
 r^2\quad {\rm for} \quad 0\leq x_i<1, \hbox{ and }
g_i(x_i)= 0\quad {\rm for} \quad -1\leq x_i\leq 0$ where $r^2$ is constant.
We find that for $r^2>-1$ the lowest frequency branch goes through origin with 
 linear behaviour, which leads to effective properties encountered in most
periodic structures. However, the case whereby
$r^2=-1$ is very unusual, as the frequency $\lambda$ behaves like $\sqrt{k}$ near the origin, where
$k$ is the wavenumber. Finally, when $r^2<-1$, the lowest branch does not pass
through the origin and a zero-frequency band gap opens up.
%, which is the hall-mark of
%metametarials supporting a host of low frequency surface plasmons.
In the last
two cases, effective medium theory breaks down even in the quasi-static limit, while
the high-frequency homogenization [Craster et al.,  Proc. Roy. Soc. Lond. A 466, 2341-2362, 2010]
neatly captures the detailed features of band diagrams.


%\end{abstract}
%\end{frontmatter}

\vskip 0.5\baselineskip

\noindent{\bf R\'esum\'e} \vskip
0.5\baselineskip \noindent {\bf Sur une classe d'\'echiquiers \`a trois phases aux propri\'et\'es effectives singuli\`eres}
Nous \'etudions le spectre de bande associ\'e aux modes
de Floquet-Bloch dans une classe d'\'echiquiers p\'eriodiques.
Sur une cellule de base $[-1,1[^2$, l'indice de r\'efraction est d\'efini par $n^2= 1+ g_1(x_1)+g_2(x_2)$
o\`u $g_i(x_i)=
 r^2\quad {\rm (une constante), pour} \quad 0\leq x_i<1, \hbox{ et }
g_i(x_i)= 0\quad {\rm pour} \quad -1\leq x_i\leq 0$.
Pour $r^2>-1$, la premi\`ere bande passe par l'origine avec un comportement lin\'eaire,
ce qui conduit \`a des propri\'et\'e effectives rencontr\'ees dans la plupart des structures p\'eriodique.
En revanche, le cas $r^2=-1$ est moins ordinaire, puisque la bande de fr\'equences acoustiques $\lambda$
se comporte comme $\sqrt{k}$ au voisinage de l'origine, avec
$k$ le nombre d'onde. Finallement, quand $r^2<-1$, la bande acoustique dispara\^it: la premi\`ere bande
ne passe plus par l'origine et une bande interdite \`a fr\'equence nulle appara\^it. Dans ces deux derniers cas de figure,
la th\'eorie des milieux effectifs ne s'applique pas, alors que la th\'eorie d'homog\'en\'eisation hautes fr\'equences
[Craster et al.,  Proc. Roy. Soc. Lond. A 466, 2341-2362, 2010] reproduit
avec pr\'ecision
les diagrammes de bandes.

\underline{\it Mots-cl\'es~:}
{Homog\'enisation~; R\'efraction n\'egative~; Bande acoustique}

%\noindent{\small{\it Mots-cl\'es~:} Mot-cl\'e1~; Mot-cl\'e2~;
%Mot-cl\'e3}}

%\selectlanguage{francais}
\section*{Version fran\c{c}aise abr\'eg\'ee}
Nous consid\'erons un \'echiquier p\'eriodique dont la
g\'eom\'etrie est d\'ecrite dans  la figure 1. Les
ondes optiques ou acoustiques qui s'y propagent sont solutions de l'\'equation
de Helmholtz (\ref{eq:helmholtz}) dans les cases homog\`enes
occup\'ees par des mat\'eriaux isotropes qui sont positifs, nuls, ou n\'egatifs.
Par ailleurs, des conditions classiques de continuit\'e de la composante tangentielle
du champ \'electromagn\'etique (ou du champ de d\'eplacement et
 de la composante normale du stress) sont impos\'ees aux interfaces
entre les diff\'erents mat\'eriaux de la cellule \'el\'ementaire et des conditions de
Floquet-Bloch sur les bords oppos\'es de celle-ci. Dans le cadre de la th\'eorie
d'homog\'en\'eisation hautes fr\'equences d\'evelopp\'ee
dans \cite{craster1}, nous nous int\'eressons plus
particuli\`erement au cas o\`u les conditions de Floquet-Bloch se r\'eduisent
\`a des conditions p\'eriodiques ou anti-p\'eriodiques, ce qui permet de
d\'eduire l'\'equation homog\'en\'eis\'ee (\ref{eq:pde_intro})
qui conduit \`a une estimation fine des courbes de dispersion \`a travers
la fr\'equence effective (\ref{eq:asymptotic_intro}), pour des nombres
d'onde au voisinage des points A, B et C de la zone de Brillouin, cf. figure 1.
Il s'av\`ere que l'on peut m\^eme d\'eduire
les relations de dispersion analytiquement dans le cas d'\'echiquiers \`a trois phases,
donn\'ees par (\ref{dispchess}) et (\ref{eq:coupled}).

Gr\^ace \`a ces approches de type homog\'en\'eisation hautes fr\'equences
\cite{craster1}, et Kr\" onig-Penney \cite{kronigpenney},
nous mettons en exergue trois types de comportements effectifs pour des
ondes se propageant dans de tels \'echiquiers aux basses fr\'equences:
ordinaire de type cristal photonique ($r^2>-1$: bande acoustique lin\'eaire \`a l'origine,
voir figure \ref{fig4})(a)), singulier de type cristal
plasmonique ($r^2=-1$: bande acoustique non-lin\'eaire \`a l'origine, voir figure \ref{fig2}(a))
et de type m\'etamat\'eriaux ($r^2<-1$: perte de bande acoustique i.e. bande
interdite \`a fr\'equence nulle, voir figure \ref{fig2}(b)).

Nous concluons notre \'etude par une application de type lentille \'echiquier
par r\'efraction n\'egative quand $r^2>-1$, avec l'image d'un point source
\`a la fr\'equence donn\'ee par l'estimation (\ref{negcool}), voir figure \ref{fig4}(b).

%\selectlanguage{english}
% main text
\section{Introduction: Beyond effective medium theory with high-frequency homogenization}
\label{}

 \label{sec:introduction}
Negative refraction is an emerging field in photonics introduced
by Victor Veselago in the late 1960's \cite{veselago},
with renewed interest following
the controversial claim of John Pendry that negative refraction makes a flat lens
with unlimited resolution possible
 \cite{pendry_prl00}. The quest for this superlens has fuelled research in
 structured photonic materials during the last decade. These composites are
known as photonic crystals when the wavelength of light
is of the same order as the typical heterogeneity size (Bragg frequency regime),
and metamaterials when the wavelength is much larger (low frequency homogenization
regime).
%lies somewhere at the interface between surface
%science, photonics and new technologies, and a comprehensive review
%of its infancy can be found in \cite{sar_rpp05}.
The propagation of waves with
anti-parallel group and phase velocities, an essential ingredient for negative refraction,
was discussed in the late 1940's by L\'eon Brillouin \cite{brillouin}.

\begin{figure}[t]
  \begin{center}
    \includegraphics[height=5cm]{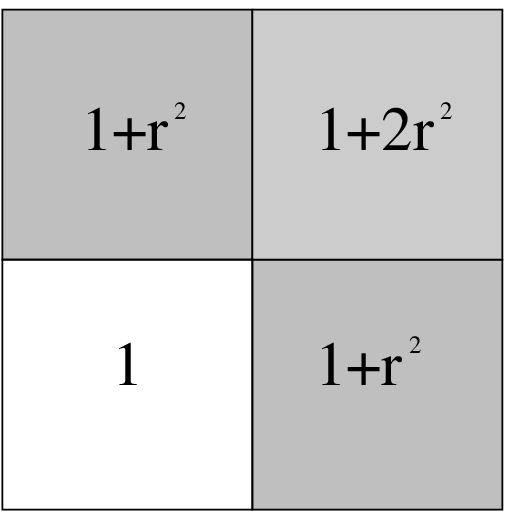}
%\end{center}
\label{fig:geometry}
%\end{figure}
    \includegraphics[height=5cm]{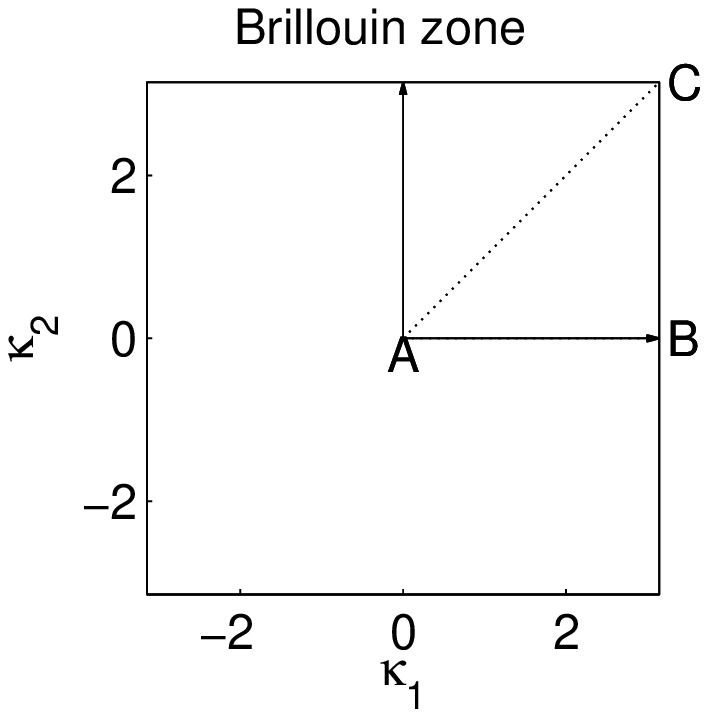}
\end{center}
\caption{Left: A single cell for the checkerboard geometry, for
  piecewise constant media.  Right: 
The reciprocal Brillouin lattice in wavenumber, $\bm\kappa=(\kappa_1,\kappa_2)$, space}
\label{fig1}
\end{figure}

In order to investigate the properties of photonic crystals,
we consider the Helmholtz equation (\ref{eq:helmholtz}) in a periodic medium.
This equation could, with appropriate notational and
linguistic changes, hold for acoustic, electromagnetic, water or
out-of-plane elastic waves and so encompasses many possible physical
applications. We solve 
\beq
 \frac{\partial^2 u}{\partial{x_1}^2}+\frac{\partial^2
   u}{\partial{x_2}^2}+ \lambda^2[ 1+ g_1(x_1)+ g_2(x_2)]u=0,
\label{eq:helmholtz}
\eeq
 for $u(x_1,x_2)$ 
 on the square $-1<x_1,x_2\leq 1$, where $\lambda^2$ is the frequency squared.
In the case of a three-phase checkerboard with a square cell as shown in Fig. \ref{fig:geometry}, $g_i(x_i)$ is taken to be 
 the piecewise constant
\beq
 g_i(x_i)=
 r^2\quad {\rm for} \quad 0\leq x_i<1, \hbox{ and }
g_i(x_i)= 0\quad {\rm for} \quad -1\leq x_i< 0 \; .
\label{eq:g}
\eeq

We note that in the context of optics, the unknown $u$ in
(\ref{eq:helmholtz}) stands for the longitudinal component of the electric field $E_z$ and
$\lambda^2$ is associated with
$\omega^2/c^2$ whereby $\omega$ is the
electromagnetic wave frequency and $c$ is the speed of light
in vacuum. Moreover, $\varepsilon_r\mu_0= 1+ g_1(x_1)+ g_2(x_2)$
where $\mu_0=1$ and $\varepsilon_r$ are the relative
permeability and permittivity of the dielectric (non-magnetic)
medium.

\begin{figure}%[h]
  \begin{center}
    \includegraphics[height=9.5cm]{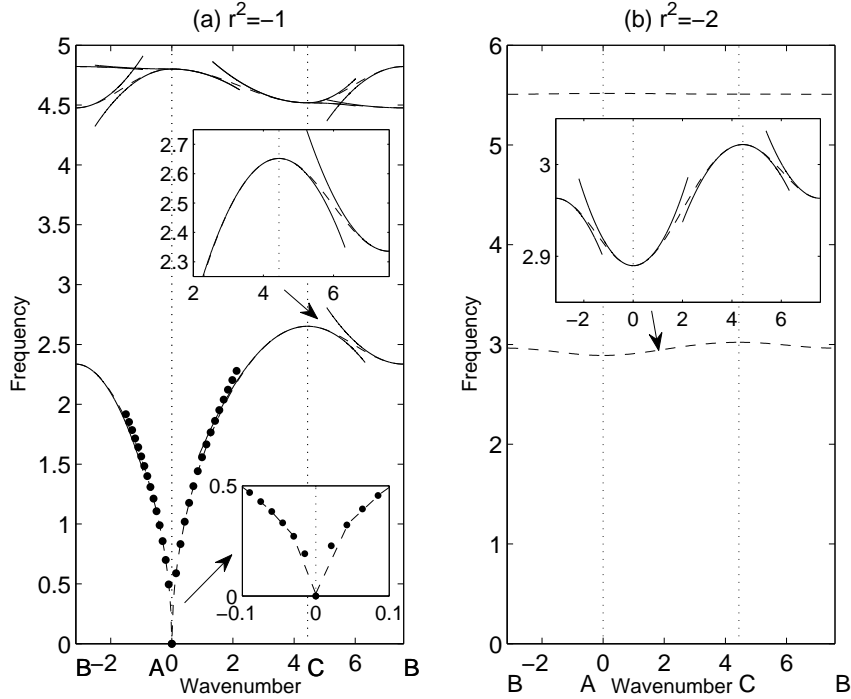}
\end{center}
\caption{Dispersion diagrams for $r^2=-1$ (a) and $r^2=-2$ (b).
Panel (a) shows the lowest three 
exact dispersion curves from the explicit dispersion relations
and the asymptotic dispersion 
curves with the behaviour at the lowest curve that touches the origin 
shown dotted. The lower inset shows the detail near the origin comparing the 
exact (dashed) and asymptotic (dotted) solutions. The upper inset shows a 
similar comparison (asymptotics given by the  solid line) for the lowest curve near 
wavenumber position C. Panel (b) shows that there is no acoustic band and that the dispersion 
curves are nearly flat. The inset shows an enhanced view of the lowest 
dispersion branch with the numerics (dashed) shown together with 
asymptotics (solid).
}
\label{fig2}
\end{figure}

For waves through an infinite, perfect, doubly periodic checkerboard one can
invoke Bloch's theorem \cite{brillouin} and simply
consider the square cell with quasi-periodic Bloch boundary conditions applied to the
edges:
\beq
%\hspace{-0.5cm}
\begin{array}{ccc}
  &u(1,x_2)=e^{\i\kappa_1}u(-1,x_2),
  u_{x_1}(1,x_2)=e^{\i\kappa_1}u_{x_1}(-1,x_2),  \nonumber \\
  &u(x_1,1)=e^{\i\kappa_2}u(x_1,-1),
  u_{x_2}(x_1,1)=e^{\i\kappa_2}u_{x_2}(x_1,-1),
\end{array}
\label{eq:bloch}
\eeq
 that involves the Bloch wave-vector ${\bm\kappa}=(\kappa_1,\kappa_2)$
 characterizing the phase-shift as one moves from one cell to the
 next. 
There is also continuity of $u, u_{x_1}, u_{x_2}$ along $x_1=0$
and $x_2=0$. This Bloch problem is solved explicitly and
dispersion relations that link the frequency and Bloch wavenumber are
deduced; as is well-known in solid state physics \cite{brillouin}
only a limited range of wavenumbers need be considered, namely
the wavenumbers along the triangle shown in the reciprocal Brillouin
lattice of Fig. \ref{fig1} and these form the irreducible
Brillouin zone. The dispersion curves reported in Fig. \ref{fig2}-\ref{fig4}
illustrate several interesting
features: stop-bands for which wave propagation is not possible, and
regions of flat dispersion curves for which the group velocity is zero
and features of slow sound or light occur. In this paper, we
focus our attention to the lowest curves, associated
with averaged properties of the checkerboards, and discuss
the extension of existing effective theories to higher frequencies.

Conventional homogenization is widely assumed
 to be ineffective for modelling photonic crystals as it is limited to
 low frequencies when the wavelength is long relative to the
 microstructural lengthscales. Here the recently developed high
 frequency homogenization theory \cite{craster1},
 which is free of the conventional limitations, is used to generate
 effective partial differential equations on a macroscale, that have
 the microscale embedded within them through averaged quantities.
We focus our attention on periodic checkerboards, some of them with sign-shifting
coefficients \cite{cras2009,bonnet2010}: the latter are known to support a host
of surface plasmons associated with highly resonant features less than conducive for
any effective theory. Our aim is actually to push
the newly introduced high-frequency homogenization beyond its limits.

At leading order, the field $u({\bf x})$ propagating inside the periodic checkerboard can be represented as
$U(\bm\xi) f({\bf X})$ where ${\bm\xi}$ and ${\bf X}$ are the short-
and long-scales \cite{craster1}. The function $U$ represents the local short-scale
solution at a specific standing wave frequency, $\lambda_0$ (which
 is in the high frequency regime), and $f$ the long-scale variation. Ultimately an
homogenized PDE emerges for $f$ entirely on the macroscale \cite{craster1}
\beq
T_{ij}\frac{\partial^2 f}{\partial X_i\partial X_j} +(\lambda^2-\lambda_0^2)
f=0,
\label{eq:pde_intro}
\eeq 
 where the spatially constant tensor $T_{ij}$ incorporates the short-scale information associated with the standing wave frequency $\lambda_0$; $\lambda$ is the full frequency. The PDE can be augmented by additional terms if there is material variation leading to localised defect modes \cite{nolde}. 
 The PDE (\ref{eq:pde_intro}) can then be utilized for, say, Bloch waves where $f\sim \exp(\i {\bf k}\cdot
{\bf X}/2)$,  
 to find local dispersion relations 
\beq
     \lambda\sim \lambda_0\left(1+\epsilon^2\frac{T_{ij}k^2_{j}}{8\lambda_0^2}\right)
\label{eq:asymptotic_intro}
\eeq
 valid near the standing wave frequency $\lambda_0$; this formula is explicitly for those periodic-periodic standing waves at wavenumber $A$ and almost identical formulae hold for the other standing waves.
Thus, the constant $T_{ij}$, once identified, completely encapsulate the effect
of the microstructure on the dispersion properties of the system; the
disparity of scales is encapsulated in $\epsilon$ which is assumed positive and 
$\ll 1$. Some typical values of the tensor $T_{ij}$ are given for the cases $r^2=-1$ and $-2$
in table \ref{tab1}. One should note that in some cases the diagonal entries, are identical
(effective isotropic parameters), different (anisotropic parameters) and even of opposite
signs (a hallmark of anomalous dispersion).

\begin{table}%[ht]
%\hspace{-1cm}
\begin{minipage}[b]{0.5\linewidth}
\centerline{$r^2=-1$}
\centering

\begin{tabular}{ccccc}\hline
  & $\lambda$ & $\mu$ & $T_{11}$ & $T_{22}$ \\
$A$ & $0$ & $0$ & $N/A$ & $N/A$\\
$A$ & $4.801$ & $2.605 $ & $-2.907$ & $0.195$ \\
$A$ & $4.801$ & $2.605 $i & $0.195$ & $-2.907$\\
$C$ & $2.652$ & $0$ & $-1.813$ & $-1.813$\\
$C$ & $4.519$ & $2.345$ & $2.951$ & $-0.380$\\
$C$ & $4.519$ & $2.345$i & $-0.380$ &  $2.951$ \\
$B$ & $2.336$ & $0.767$i & $-3.445$ & $1.394$\\
$B$ & $4.477$ & $2.350$i & $3.215$ & $0.384$\\
$B$ & $4.821$ & $2.595$ & $-0.199$ & $-2.774$\\
\hline
\end{tabular}

\end{minipage}
%\hspace{-4cm}
\begin{minipage}[b]{0.5\linewidth}
\centerline{$r^2=-0.9$}
\centering
\begin{tabular}{ccccc}\hline
  & $\lambda$ & $\mu$ & $T_{11}$ & $T_{22}$ \\
$A$ & $0$ & $0$ & $N/A$ & $N/A$\\
$A$ & $4.678$ & $2.544 $ & $-4.436$ & $0.278$ \\
$A$ & $4.678$ & $2.544 $i & $0.278$ & $-4.436$\\
$C$ & $2.6086$ & $0$ & $-2.1778$ & $-2.1778$\\
$C$ & $4.263$ & $2.1596$ & $4.079$ & $-0.678$\\
$C$ & $4.263$ & $2.1596$i & $-0.678$ &  $4.079$ \\
$B$ & $2.2473$ & $0.8165$i & $-4.3226$ & $1.4607$\\
$B$ & $4.1814$ & $2.1623$i & $4.7095$ & $0.6936$\\
$B$ & $4.708$ & $2.530$ & $-0.2883$ & $-4.1548$\\
\hline
\end{tabular}
\end{minipage}

\caption{The standing wave frequencies, $\lambda$, the $\mu$ and the corresponding $T_{11}$ and $T_{22}$ for the lowest three dispersion curves for $r^2=-1$ and $r^2=-0.9$}
\end{table}

\begin{table}
\centering

\begin{tabular}{ccccc}\hline
  & $\lambda$ & $\mu$ & $T_{11}$ & $T_{22}$ \\
$A$ & $2.8898$ & $0$ & $0.4519$ & $0.4519$\\
$C$ & $3.0217$ & $0$ & $-0.3412$ & $-0.3412$\\
$B$ & $2.9632$ & $0.3524$ & $-0.41$ & $0.3596$\\
\hline
\end{tabular}

\caption{Values of $\lambda$ and the corresponding $\mu$ at the  wavenumbers for standing waves with the associated values of $T_{11}$ and $T_{22}$ for the lowest dispersion curve for $r^2=-2$.}
\label{tab1}
\end{table}

\section{An exact dispersion relation for a class of three-phase checkerboards}
We have first investigated the dispersion curves associated with three-phase
periodic checkerboards using the finite element
software COMSOL MULTIPHYSICS. However, the convergence of the
numerical algorithm can be hazardous, due to sign-shifting coefficients \cite{bonnet2010}.
Thankfully, the Bloch problem for such checkerboards can be solved analytically and this
provides probably the only non-trivial two dimensional
structure with an exact solution; it is a natural generalization
of the classical Kronig-Penney \cite{kronigpenney}
one-dimensional
piecewise constant periodic medium.
The dispersion relation follows from setting $u(x_1,x_2)=Q_1(x_1)Q_2(x_2)$ in the
governing equation (\ref{eq:helmholtz}). 
From the piecewise constant fonction (\ref{eq:g}), and
from the Bloch conditions (\ref{eq:bloch}), the following coupled  equations emerge \cite{craster3}
\beq
 2\gamma_i\beta_i(\cos\kappa_i-\cos\gamma_i\cos\beta_i)+(\beta_i^2+\gamma_i^2)\sin\gamma_i \sin\beta_i=0,
\label{eq:coupled}
\eeq
 for $i=1,2$. These are dispersion relations for $\lambda^2,\mu^2$ (the latter is a separation constant) in
 terms of $\kappa_1,\kappa_2$; the parameters $\gamma_i,\beta_i$ act
 to couple the equations as
\beq
   \beta_i^2={\lambda^2/2}\mp \mu^2,\qquad
   \gamma_i^2=\lambda^2{(1/2+r^2)}\mp \mu^2,
\label{dispchess}
\eeq
 for $i=1,2$ with the minus, plus signs for $i=1, 2$
 respectively. This two variable dispersion relation is solved via
  matrix Newton iteration. For the exact solution found in this
  section, one finds $\lambda,\mu$ explicitly and some sample
  eigenvalues are given in table \ref{tab1}. Notably some
  frequencies, $\lambda$, have two associated $\mu$ values, one real
  and one imaginary and these correspond to degenerate cases where the dispersion curves touch.

\section{Illustrative electromagnetic paradigms whereby effective medium theory fails}
Effective properties of periodic checkerboards are of particular mathematical interest,
and their study goes back a long way, with the seminal paper by Keller \cite{keller}
for two-phase checkerboards, and those of Mortola-Steffe \cite{mortola}, Craster-Obnosov
\cite{craster2} and Milton \cite{milton} for four-phase checkerboards.  
Physicists have further analysed the effective properties of two-phase checkerboards with continuously varying
parameters in the context of photonic crystals \cite{pre2003}.
In this note, we investigate the case of three-phase checkerboards with negative and vanishing
refractive index (when $r^2$ takes negative values).
Such a study is motivated by the fabrication of nano-scale gold and silver checkerboards whose cells
alternate positive and negative refractive index media in the visible range of frequencies. Potential
applications lie in extra-ordinary transmission of light through the sub-wavelength aperture holes
in such checkerboards, and tremendously enhanced local density of states for light confinement \cite{cras2009}.
These physical phenomena are underpinned by the effective properties of such checkerboards. Hence, an adequate
homogenization model for such structures is of pressing importance.

We start to explore the case of a checkerboard with $r^2=-1$. Dispersion diagrams reported in 
figure \ref{fig2}(a) compare the solution from the exact dispersion relations with those from the asymptotics. It clearly demonstrates the
superiority of the high-frequency homogenization approach which reproduces very precisely the acoustic
and optical branches, thereby extending classical homogenization to the stop band (Bragg) regime of frequencies. Moreover, the region where the lowest branch cuts the origin, the long wave low frequency regime, is no longer linear and asymptotics of the dispersion relation show that the local behaviour is that $\lambda\sim 6^{1/4}\vert\kappa\vert^{1/2}$. %(c.f figure  \ref{fig4}).
 The insets show detail of the asymptotics versus the exact solution.

Figure \ref{fig2}(b) shows the case of a checkerboard with $r^2=-2$; the
acoustic band is lost for $r^2<-1$, which is reminiscent of singular problems in homogenization of periodic arrays of
infinitely conducting fibres in transverse electric polarization (modelled with Dirichlet boundary conditions)
\cite{sasha}. In the present case, we do not consider Dirichlet conditions (corresponding to hard walls in acoustics)
 but we rather have sign-shifting conditions across some interfaces. It is clear that effective medium theory breaks
down, however the high-frequency homogenization approach captures the fine
features of the nearly flat dispersion curves, see the inset to figure \ref{fig2}(b), which are associated with slow waves,
another topical subject in photonics.

\begin{figure}[h]
  \begin{center}
    \includegraphics[height=5.5cm]{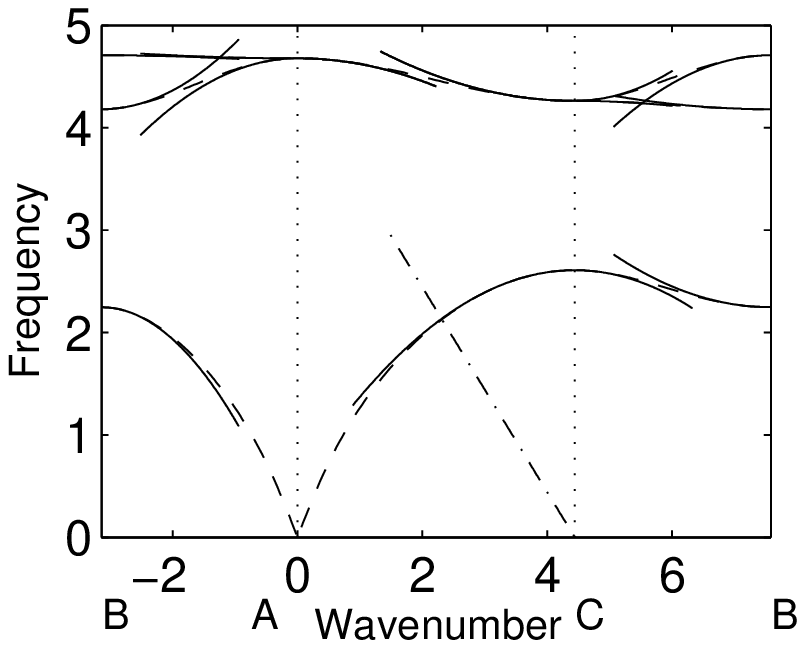}
    \includegraphics[height=5.5cm]{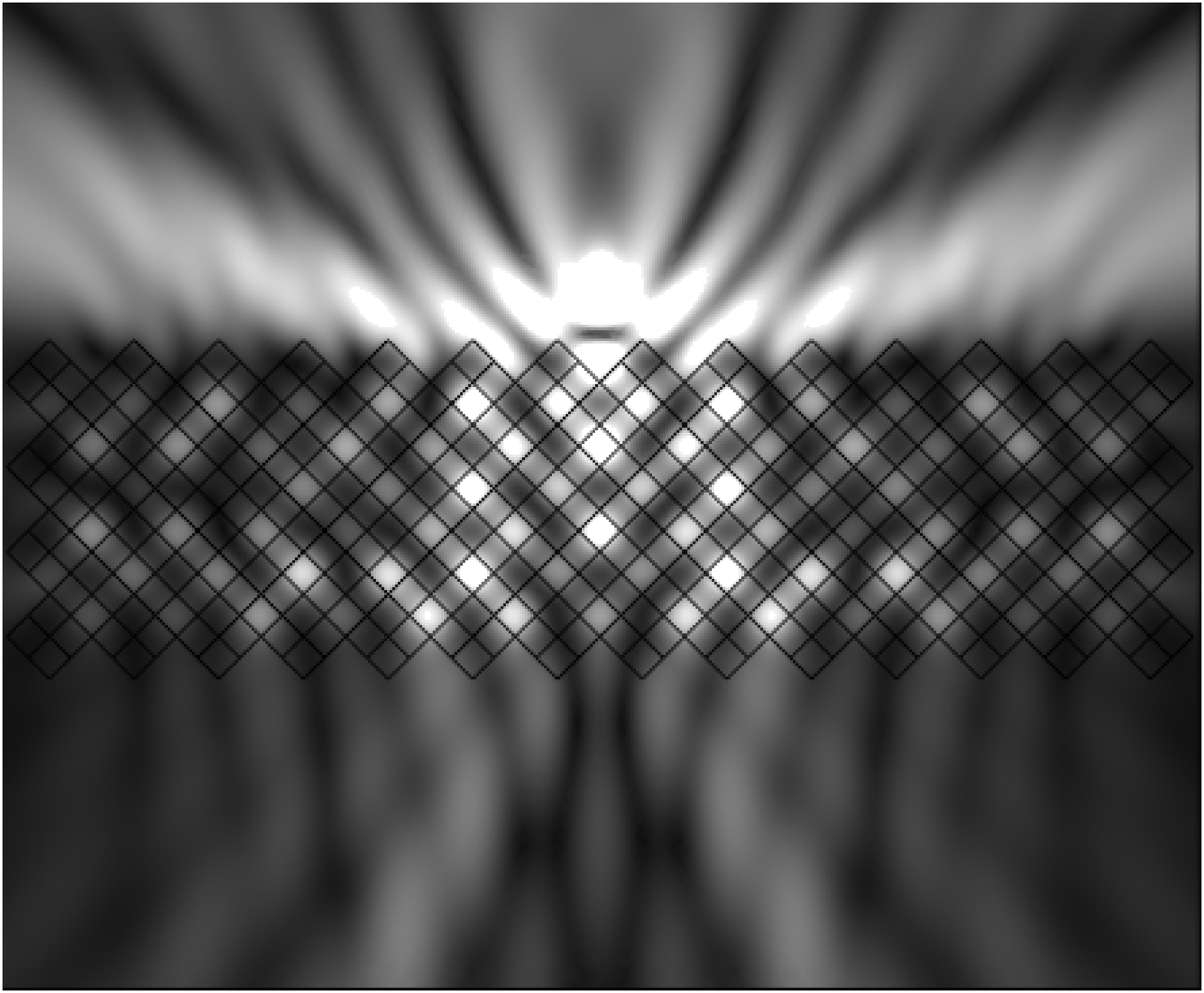}
\end{center}
\caption{Dispersion diagrams for $r^2=-0.9$. Left panel shows the dispersion 
curves that are visually close to those for $r^2=-1$ however, 
 the behaviour near $A$ for low frequency is 
asymptotically different between these cases. The dot-dash line is the light-line for waves impinging upon the rotated checkerboard. 
 The right panel shows the response due to a harmonic point source at frequency $\lambda=2.1$ above a finite size checkerboard for
$r^2=-0.9$ with
each cell rotated by an angle $\pi/4$; it displays an image below the checkerboard in accordance
with the Snell-Descartes law for a negative refractive index.}
%The inset shows the lowest dispersion curve 
%along AC on log-log axes with the dashed line for $r^2=-1$ for which the slope 
%is $1/2$, the barely discernible dotted line is $\lambda \sim 
%6^{1/4}\sqrt{\vert{\bf \kappa}\vert}$. The solid line is for 
%$r^2=-0.9$ showing the usual $\lambda\sim\vert\kappa\vert$ linear 
%behaviour.
\label{fig4}
\end{figure}

We finally study the case of a checkerboard with $r^2=-0.9$, figure \ref{fig4}(a) looks visually
similar to figure \ref{fig2}(a); however, a closer view on a log-log scale (not shown) of the acoustic band reveals that the long wave low frequency limit differs from that when $r^2=-1$. For $r^2=-0.9$ one gets the usual effective medium result with frequency linearly related to wavenumber however for $r^2=-1$ 
the effective medium theory no longer holds and a different asymptotic relation holds ($\lambda\sim 6^{1/4}\vert\kappa\vert^{1/2}$); we further observe that the high-frequency homogenization approach (dotted line) captures the fine features of this curve (dashed line).

Figure \ref{fig4}(a) also shows the light-line emerging from $C$ relevant for waves impinging upon a checkerboard structure rotated through $\pi/4$. Notably the light-line crosses the lowest dispersion curve in such a way that their group velocities are opposite and therefore one can induce negative refraction at this frequency. The asymptotic theory explicitly identifies this frequency as
\beq
\lambda=\lambda_0/\sqrt{1-T/4}
\label{negcool}
\eeq
 where $T_{11}=T_{22}=T$ for that case. The estimate is that $\lambda\sim 2.1$ and computations for this are shown in figure \ref{fig4}.

\vspace{-1cm}

\section{Conclusions}
We have investigated the Bloch spectrum of periodic, three-phase,
checkerboards. The focus has been on special cases in which
the refractive index squared (i.e. the relative permittivity in the
contex of optics) in homogeneous cells can take vanishing and/or negative values,
and we have found that anomalous dispersion effects of
waves are possible, including acoustic bands with non linear features
for vanishing wavenumbers, zero frequency stop bands, and nearly flat
dispersion curves. The last two items are known to be respectively
associated with singularly perturbed problems in electromagnetism that
cannot be homogenized in a conventional way \cite{sasha}, as no
wave is allowed to propagate in the periodic structure at low frequencies, 
and slow waves, whereby delay lines can be achieved.  

% etc, etc

% \appendix

% \section{}
% \label{}

% The Acknowledgements are also a un-numbered section
%\section*{Acknowledgements}
% Acknowledgements text here

%% \bibliography{references}
%% \bibliographystyle{newJFM}

%% \begin{thebibliography}{00}
%% % please try to use the bibitem system -
%% %   the references should be in order of citation in the text
%% % \bibitem{label}
%% % Text of bibliographic item
%% \end{thebibliography}

\end{document}